\documentclass{article}
\usepackage{spconf,amsmath,graphicx,arabtex,utf8,tabularx,booktabs,url}
\usepackage{subcaption}
\usepackage[multiple]{footmisc}
\usepackage[table]{xcolor}
\usepackage{CJKutf8}
\usepackage{enumitem}
\ninept

\title{Benchmarking Evaluation Metrics for \\ 
Code-Switching Automatic Speech Recognition}

\name{\begin{tabular}{c}Injy Hamed,$^{1,2}$ Amir Hussein,$^{3}$ Oumnia Chellah,$^4$ Shammur Chowdhury,$^5$ \\Hamdy Mubarak,$^5$ Sunayana Sitaram,$^6$ Nizar Habash,$^1$ Ahmed Ali$^5$\end{tabular}}
\address{$^1$New York University Abu Dhabi $^2$University of Stuttgart $^3$Johns Hopkins University\\
$^4$Stanford University $^5$Qatar Computing Research Institute $^6$ Microsoft Research India\\
{\tt injy.hamed@nyu.edu, ahussei6@jhu.edu}\\}
\copyrightnotice{978-1-6654-7189-3/22/\$31.00~\copyright2023 IEEE}
\begin{document}
%
\maketitle
\begin{abstract}


Code-switching poses a number of challenges and opportunities for multilingual automatic speech recognition. In this paper, we focus on the question of robust and fair evaluation metrics. To that end, we develop a reference benchmark data set of code-switching speech recognition hypotheses with human judgments. We define clear guidelines for minimal editing of automatic hypotheses. We validate the guidelines using 4-way inter-annotator agreement. We evaluate a large number of metrics in terms of correlation with human judgments.  The metrics we consider vary in terms of representation (orthographic, phonological, semantic), directness (intrinsic vs extrinsic), granularity (e.g. word, character), and similarity computation method. The highest correlation to human judgment is achieved using transliteration followed by text normalization. We release the first corpus for human acceptance of code-switching speech recognition results in dialectal Arabic/English conversation speech.

\end{abstract}
\begin{keywords}
ASR, Code-switching, Evaluation metric
\end{keywords}

\setcode{utf8}

\section{Introduction} 


Code-switching (CS) is the act of using more than one language within the same discourse. The prevalence of CS across multi-cultural and multi-lingual societies has been met with a growing interest in the NLP and speech processing fields. Automatic Speech Recognition (ASR) for CS is a well studied problem \cite{SCR+19} conducted in several language pairs on acoustic modeling, language modeling and novel system architectures. 

In code-switched language pairs with different scripts, there is a tendency to cross-transcribe words that creates a large number of homophones in the data, leading to challenges in training and evaluation. 
Rendering the CS transcription accurately is important, however, is often not straight-forward, and can be inconsistent, leading to the same word being transcribed using different scripts \cite{SS18}. Cross-transcription is particularly challenging in languages having a high amount of loanwords, as it is not always clear which language a word belongs to, and hence, which script to transcribe it in. 

\begin{table*}[h]
\centering
 \includegraphics[width=0.9\textwidth]{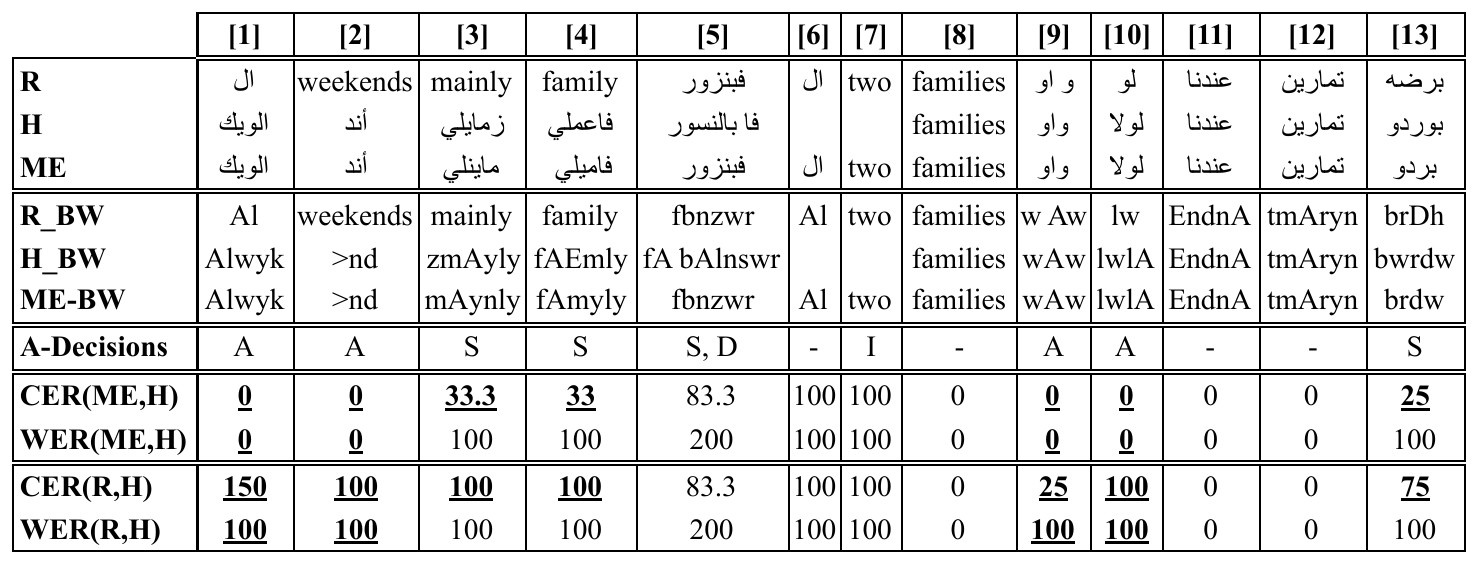}
 \caption{This example shows a hypothesis (H), the minimal edits annotation (ME), the ArzEn transcription reference (R), and their Buckwalter transliterations \cite{Habash:2007:arabic-transliteration} (R\_BW, H\_BW, and ME\_BW). We denote annotation decisions (A-Decision) as substitution (S), insertion (I), deletion (D), or acceptance (A) for words having different forms across H and ME. 
 In this example, CER and WER reductions are seen for minimal edits annotations over reference due to cross-transcription (segments 1, 2, 3, and 4) and unstandardized orthography (segments 9 and 13) issues.
 }
 \label{table:fun_example}
\end{table*}

Investigating the ideal ASR performance metric for CS, \cite{SS18} propose a modified WER metric based on mapping both languages into the pronunciation space, which leads to improvements in accuracy and evaluation. \cite {ERR+18} propose a transliteration-based WER metric by transliterating mixed-script utterances into a single script. The authors demonstrate the robustness of the proposed approach on several Indic languages. However, these techniques can also lead to false positives when there are words in the two languages that sound the same but have a different meaning. It also has limitations in cases where different parts of the word are written in different scripts. e.g., the word \textbf{\<آرت>ficial}, which is \textbf{artificial} in English script or \textbf{\<آرتفيشيال> } in Arabic script. 
Thanks to word-pieces \cite{chowdhury2021towards}, we may expect partial (cross-script) transcription in CS ASR results. 

Researchers have also investigated techniques for handling the problem of non-standardized orthography for Dialectal Arabic ASR evaluation. 
\cite{AMR15} proposed the use of Multi-Reference Word Error Rate to allow for a wider coverage of different spelling variants for Dialectal Arabic ASR evaluation. 
Whereas, \cite{AKH19} investigated approaches to reduce spelling variations, which included normalization and following the Conventional Orthography for Dialectal Arabic (CODA) \cite{HDR12} guidelines to spell words, as well as relying on morphologically abstracted forms obtained from different tokenization schemes and lemmatization. 



This work builds upon previous contributions from \cite{ERR+18} and \cite{SS18}. In this study, we investigate various methods that go beyond orthographic transliteration methods. 
We explore lexical and phonetic representations for evaluation, and study various weighted edit-distance methods and semantic similarity evaluation. 
We designed a guideline and developed a reference human acceptability corpus (HAC) -- quantifying the human judgments in terms of minimal editing of ASR hypothesis. We evaluate a large number of metrics in correlation with human judgments. 
We believe that this is the first study on human acceptability for CS speech recognition. 

\noindent The contributions of this paper are as follows:
\setlength\itemsep{-0.1em}
\setlist[itemize]{noitemsep, topsep=-0.1pt}
\begin{itemize} 

\item We design and develop the first corpus for human acceptance for CS speech recognition. The corpus and its guidelines are publicly available.\footnote{\url{http://arzen.camel-lab.com/}}

\item We introduce phone similarity edit distance (PSD) and show its correlation with human acceptance.
\item We propose a novel approach to use machine translation on hypotheses and references and report results using semantic evaluation to overcome the cross-transcription challenge.
\end{itemize}


\section{HAC: Human Acceptability Corpus for Code-Switching}  

%
%
%
%
%
%
%
%
%


\vspace{-0.1mm}
For developing the Human Acceptability Corpus for Code-switching (HAC), we use a subset from the ArzEn Egyptian Arabic-English CS conversational speech corpus \cite{HVA20} and obtain the hypotheses using different ASR systems trained on publicly available corpora. 
We carefully design the human annotation task to indicate the amount of post-editing effort needed to correct the hypotheses. While previous work has relied on human judgments in the form of systems' ranking \cite{AKH19,KLZ+21}, we opt for a more fine-grained human evaluation, where each hypothesis is evaluated independently in terms of the number of edits performed by the annotators. Throughout the paper, we will refer to the annotators' post-edited text as `minimal edits annotations' and to the original ArzEn transcriptions as `references'.

\subsection{Annotation Guidelines}
The annotators were provided with an audio file for each utterance 
and its hypothesis. 
The references were not provided to avoid biasing the annotations. The annotators were asked to perform minimal edits to make the hypotheses acceptable, obeying the following rules:
\begin{itemize}
    \item \textbf{Rule of Script Segregation:} Words should be written in Arabic or Roman script, and not a mix of the two. The only exception is the writing of Arabic affixes and clitics in conjunction with English words. Arabic words should be in Arabic script. English words can be in Arabic or Roman script (default is Roman). For missing words, the default script of the language should be used. 
    English origin words that have been integrated in Arabic templatic morphology, and phonology will be treated as Arabic words. These guidelines differ from the guidelines used by \cite{HVA20} for collecting ArzEn transcriptions, where the transcribers were suggested to use Arabic script for Arabic words and Roman script for English words.
    \item \textbf{Rule of Acceptable Readability:} The transcript should be made readable enough to allow someone to reproduce the original audio and intended meaning. 
    \item \textbf{Rule of Minimal Edit:} Spelling variations are acceptable as long as they do not break the rule of acceptable readability. This rule covers the cases where certain letters can be used interchangeably, such as  
    \<ت>-\<ث>,
    \<ق>-\<ئ>,
    \<ي>-\<ى>,
    \<ا>-\<إ>-\<أ>, and 
    \<ة>-\<ه>. Non-verbal speech effects should not be added if absent from the ASR output. If they are included and in the audio, they should be accepted. If they are not in the audio, they should be deleted. 
\end{itemize}

In Table \ref{table:fun_example}, we provide an example demonstrating the minimal edits annotation. While the CER and WER for the hypothesis and reference are 50.0\% and 78.6\%, the error rates are dropped to 21.6\% and 46.2\% when comparing minimal edits annotation and hypothesis, showing the high amount of characters and words mispenalized by CER and WER.

\subsection{Design Consideration}
We sampled two hours of speech from ArzEn training set, consisting of seven recordings and covering a total of 1,304 utterances. For each utterance, we obtained the ASR hypotheses from three different ASR systems, resulting in a total of 3,903 hypotheses to be annotated for minimal correction.\footnote{Three utterances were excluded from the dataset as they only contain non-speech tags, resulting in 1,301 utterances, for each of which we have three ASR hypotheses.} These hypotheses were annotated by four Arabic-English bilingual speakers. 
We use three pretrained bilingual (Arabic-English) ASR systems, to generate the hypotheses. The systems vary either in architecture or in the decoding parameters, as described below. 

\paragraph*{HMM-DNN:} We used a grapheme-based model trained using a Time Delay Neural Network (TDNN) \cite{peddinti2015time} with the LF-MMI objective \cite{povey2016purely}. For training the model, we used the alignments from the context-dependent Gaussian Mixture Model and Hidden Markov model (GMM-HMM) system. For decoding, we opt for the $4$-gram model, trained on the ASR transcription. 

\paragraph*{End-to-End ASR:} For the end-to-end (E2E) ASR \cite{chowdhury2021towards} system we used a transformer based architecture~\cite{vaswani2017attention}, comprised of two sub-networks: conformer encoders and the transformer decoders \cite{conformer}. The ASR system consists of 12 encoder layers and 6 decoder layers, each with 2,048 encoder/decoder units from the feed-forward layers, and 8 attention heads with 512 transformation dimensions. 
Note that E2E model uses word-piece byte-pair-encoding (BPE)~\cite{kudo2018sentencepiece}, with size of $\sim10k$.
For the study, we adopted two variations of this model, by changing the size of the beam search in the decoding space. The variants are: \textit{(1)} Conformer-Accurate (\textit{Conformer-A}) -- with a beam size of $60$ and \textit{(2)} Conformer-Fast (\textit{Conformer-F}) with a beam size of $2$.



\subsection{Inter-annotator Agreement}
\begin{table}[t]
\centering
\setlength{\tabcolsep}{2pt}
\begin{tabular}{|l|rrr|r|r|}
\cline{2-6}
\multicolumn{1}{c|}{} & \multicolumn{1}{c}{\textbf{A2}} & \multicolumn{1}{c}{\textbf{A3}} & \multicolumn{1}{c|}{\textbf{A4}} & \multicolumn{1}{c|}{\textbf{H}} & \multicolumn{1}{c|}{\textbf{R}}\\\hline
\textbf{A1}  & \textit{8.3} / \underline{16.7}  & \textit{6.3} / \underline{15.0}  & 8.3 / \underline{18.4}  & \textit{14.9} / \underline{27.8} & \textit{19.9} / \underline{44.0}\\
\textbf{A2}  &\cellcolor{gray!25}& \textit{8.9} / \underline{17.6}  & \textit{10.8} / \underline{20.7} & \textit{14.1} / \underline{24.6} & \textit{20.5} / \underline{44.9}\\
\textbf{A3}  &\cellcolor{gray!25}&\cellcolor{gray!25}& \textit{6.9} / \underline{15.8}  & \textit{15.1} / \underline{27.9} & \textit{18.6} / \underline{41.4}\\
\textbf{A4}  &\cellcolor{gray!25}&\cellcolor{gray!25}&\cellcolor{gray!25}& \textit{16.0} / \underline{29.0} & \textit{20.5} / \underline{44.9}\\\hline
\textbf{Avg}&&& \textit{8.2} / \underline{17.4}  & \textit{15.0} / \underline{27.3} & \textit{19.9} / \underline{43.8}\\\hline        
\end{tabular}
\caption{Inter-annotator agreement, showing \textit{CER} / \underline{WER} between annotators' minimal edits, as well as each annotator with ASR hypotheses (H) and ArzEn references (R).}
\label{table:IAA}
\vspace{-0.2cm}
\end{table}

We randomly sampled 203 sentences, annotated by the four annotators, for inter-annotator agreement (IAA).\footnote{Two sentences were excluded from the IAA calculations as they were annotated as unclear.} 
The IAA evaluation is presented in Table \ref{table:IAA},  where we report the CER/WER between every two annotators. As shown, on average, the CER and WER measured between annotators are 8.2\% and 17.4\% respectively. While these figures are relatively high, it is not surprising, and reflects the complexity of the task, where word acceptability and choice may differ across annotators due to unstandardized orthography.  
We present the CER/WER between minimal edits annotations and hypotheses reflecting the amount of edits performed by the annotators. Moreover, we show the CER/WER between minimal edits annotations and references reflecting the amount of characters/words that would be mispenalized when evaluating minimal edits annotations against references using CER/WER. These numbers are also a good indicator of the limitation of CER and WER as accurate evaluation metrics. 

\section{Metrics Under Evaluation} 
Using the developed corpus, we assess multiple evaluation metrics against the ground truth error, measured in terms of human post-editing effort. The metrics we consider vary in terms of  representation (orthographic, phonological, semantic), directness (intrinsic vs extrinsic), granularity (word vs character), and similarity computation method. 


\subsection{Orthographic Metrics}
In the orthographic space, we investigate the use of four performance measures: WER, CER, Match Error Rate (MER), and Word Information Lost (WIL) \cite{MMG04}.
  \begin{align}
    &WER = \frac{S+D+I}{H+S+D}\\
    &MER = \frac{S+D+I}{H+S+D+I}\\
    &WIL=1-\frac{H^2}{(H+S+D)(H+S+I)}
  \end{align}

\noindent where \textit{H}, \textit{S}, \textit{D} and \textit{I} correspond to the number of word hits, substitutions, deletions and insertions. 
MER computes the probability of a given match between reference and hypothesis being incorrect. By including S, D and I in the denominator, the range of MER has the bound [0,1]. 
WIL, introduced in \cite{Mor02}, is an approximate measure reflecting the proportion of words lost between hypothesis and reference. 

While these measures vary in granularity, they all fall short against the cross-transcription issue. 
Therefore, following the work of \cite{SS18}, we investigate the effect of transliterating hypotheses and references into primary as well as secondary language scripts on alleviating the cross-transcription problem.
We perform automatic transliteration using the transliteration API provided by QCRI \cite{DSH+14}\footnote{\url{https://transliterate.qcri.org/api}}. In Table 
\ref{table:transliterationPSDexample}, we present an example showing the reduction achieved in CER and WER by mapping the texts into one common script, handling both cross-transcription (columns 1-3) and orthography unstandardization (column 4) challenges. One drawback of this technique though is its dependence on the availability of a language-specific transliteration system, and that its effectiveness is tied to the performance of that system. 
Following the work of \cite{AKH19}, in order to reduce the spelling variation resulting from dialectal Arabic unstandardized orthography, we investigate applying Alif/Ya normalization as a variant for the experiments relying on orthographic evaluation metrics.

\begin{table}[t]
\centering
 \includegraphics[width=0.45\textwidth]{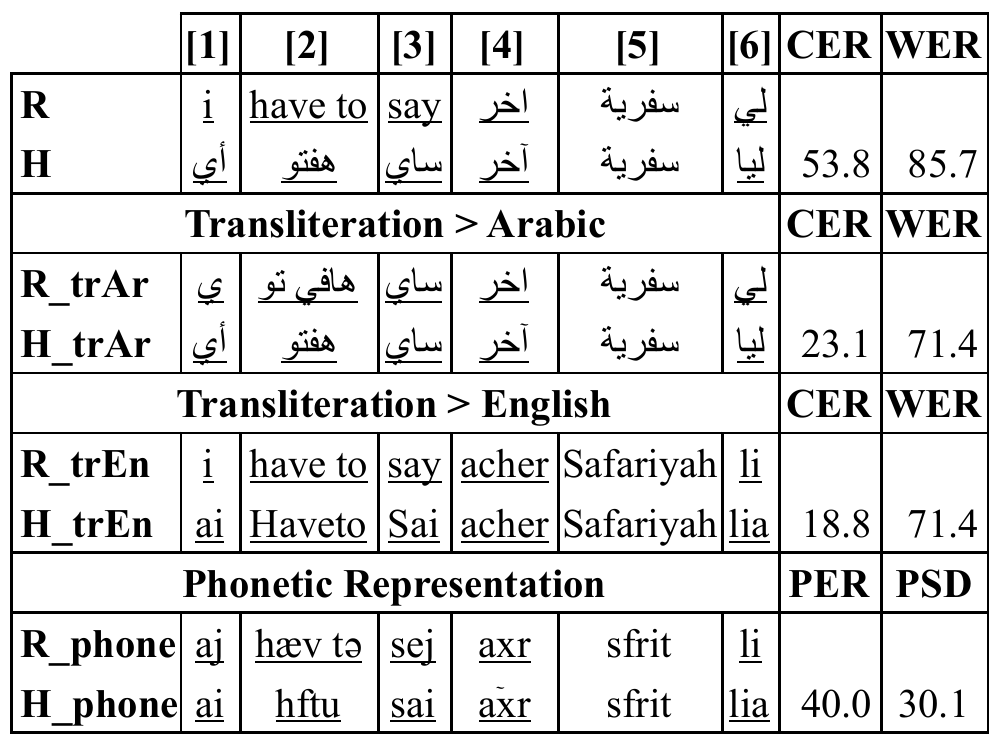}
 \caption{Example for a reference (R) and hypothesis (H), showing their corresponding transliteration output to Arabic (R\_trAr and H\_trAr) and Roman (R\_trEn and H\_trEn) scripts, as well as their IPA phone mapping (R\_phone and H\_phone) using Epitran. In this example, we separate the phones at word boundaries for better readability, however, the spaces are not included in PER and PSD calculations.
 }
 \label{table:transliterationPSDexample}
\end{table}
\begin{table*}[th!]
\centering
\begin{tabular}{lrc}
\hline
&\multicolumn{1}{c}{\textbf{Text}}& \multicolumn{1}{c}{\textbf{CosineSim}} \\\hline
\textbf{R}&
\<عادي وكان كلهم عربي>
\underline{\<\textbf{ناشيونال سكولز}>}
\<كان تعليمي لغاية الجامعة كان في >
&\\
\textbf{H}
&
\textbf{\<عادي و كان كلهم عربي>
\underline{\textbf{national schools}}}
\<كان تعليمي لغاية الجامعة كان في>
&
0.906\\\hline
\textbf{R$_{En}$}&My education until university was in normal \underline{\textbf{national schools}}, and they were all Arab &\\
\textbf{H$_{En}$}&My education until university was in a normal \underline{\textbf{National School}}, and they were all Arab&0.942\\\hline
\end{tabular}
\caption{Example presenting a reference (R) and hypothesis (H) along with their translations (R$_{En}$) and (H$_{En}$), showing how the cross-transcription issue was resolved through translation. The cosine similarities between R-H and R$_{En}$-H$_{En}$ are shown, where the embeddings are obtained from mBERT.
}
\label{table:semSim_example}
\end{table*}

\subsection{Phonological Metrics}


The orthographic-based metrics only consider literal correction, and do not adequately handle orthographic unstandardization. 
To overcome such limitation, we propose phone similarity edit distance (PSD),\footnote{\url{https://github.com/JSALT2022CodeSwitchingASR/Evaluation}} where we measure the edit distance between hypotheses and references in the shared phone space using International Phonetic Alphabet (IPA) mapping. The main advantage of this approach over the transliteration is that the IPA mapping is model independent and deterministic based on the grapheme to phoneme (G2P) dictionary. The IPA uses standardized representation of speech sounds across different languages. When calculating PSD, we scale the substitution cost by the dissimilarity between the phones based on the articulation features. As a result, PSD adds partial substitution penalty when the phones are different but close in pronunciation. We use the Epitran toolkit \cite{Mortensen-et-al} that contains massive multilingual G2P including Arabic (ara-Arab) and English (eng-Latn) languages. PSD is calculated as:
\begin{equation}
    PSD = \frac{w_S\sum{S_i} + w_D\sum{D_j} + w_I\sum{I_k}}{N}
\end{equation}
\noindent where $S$, $D$, $I$ are number of phones substitutions, deletions, and insertions, respectively; $N$ is the number of phones in the reference; $S_i = 1-sim(x_i,y_i)$ where x and y are the aligned phones. The $w_S, w_D, w_I$ are the costs for substitution, deletion and insertion respectively with value of $1$ by default. We investigate different values for $w_S$ (${1,2,4,8}$), while keeping $w_D=1, w_I=1$. 
In Table 
\ref{table:transliterationPSDexample}, 
it can be seen that our PSD implementation can easily handle  
code switching. 
However, the main shortcoming of PSD approach is that it does not consider the semantics of the words.

\subsection{Semantic Metrics}

Next, we assess the ASR output considering semantic similarity. 
Following \cite{KAD+21}, we measure the semantic similarity between reference and hypothesis pairs as the cosine similarity between their embeddings obtained from pretrained transformer models. The embeddings are generated using mean pooling over token embeddings.\footnote{We also investigate the use of CLS token, however the correlations are significantly lower.} 
While this approach has been previously investigated for monolingual ASR evaluation \cite{KAD+21,KLZ+21}, it has not been investigated in the scope of CS. 

We also introduce a novel pipeline for semantic-based ASR evaluation, where we translate the hypotheses and references into monolingual sentences using Google Translate API\footnote{\url{https://cloud.google.com/translate}}. The translations, as well as the original reference-hypothesis pairs, are then evaluated in terms of the following machine translation (MT) evaluation metrics: BLEU \cite{PRT+02}, chrF 
\cite{Pop17}, and BertScore (F1)
\cite{ZKW+19}, in addition to cosine similarity. 
We explore translating the sentences into the primary language (Arabic), secondary language (English), as well as a completely independent language (Japanese), to investigate the ability of the approach to generalize on different languages. 
For calculating cosine similarity,  we explore the use of several pretrained models, including mBERT and language-specific BERT models, for obtaining the sentence embeddings. For the original texts containing CS, we use mBERT \cite{DCL+18} and BiBERT \cite{XVM21}. For Arabic translations, we use mBERT, CAMeLBERT \cite{IAB+21}, and AraBERT \cite{ABH20}. For English translations, we use mBERT and BERT-base \cite{DCL+18}. For Japanese translations, we use mBERT, BERT-base-japanese\footnote{\url{https://huggingface.co/cl-tohoku/bert-base-japanese}}, and BERT-large-japanese\footnote{\url{https://huggingface.co/cl-tohoku/bert-large-japanese}}.\footnote{We only present the results for the best settings. We show the results with lowercasing and performing Alif/Ya normalization, which have shown to improve correlations. We do not present results for chrF++ as it gave slightly lower correlations compared to chrF. For monolingual English translations, BertScore(F1) using mBERT gave higher correlations than BertScore(F1) using roberta-large. For cosine similarity, using bert-base-multilingual-cased has also shown to give overall higher correlations over bert-base-multilingual-uncased.}

This approach provides the following advantages: (1) through translation, words in different scripts or with spelling variations can be mapped to the same/similar word(s), and (2) by using semantic similarity, such words can be assigned lower errors if closely represented in the embedding space. 
As seen in Table \ref{table:semSim_example}, the words `\<ناشيونال سكولز>' were successfully mapped through translation to `National Schools'. 
One limitation of this approach, however, is that it is highly dependent on the quality of the embeddings obtained from the pretrained models as well as MT performance.

\section{Experimental Results} 
\subsection{Experimental Setup}

We define the ground truth error for each hypothesis to be the amount of human post-editing effort required to correct it. Accordingly, we define $GoldCER$ to be the edit distance between the hypothesis and minimal edits annotation calculated using CER. 
We opt for using CER in this calculation over other evaluation metrics as it provides higher granularity in reflecting the effort done by annotators and is consistent with the annotation guidelines.

In order to assess the performance of the evaluation metrics, we compare their scores against $GoldCER$ on both the \textit{sentence-level} and \textit{system-level}.\footnote{We exclude 17 utterances that are annotated as unclear.} In the sentence-level evaluation, we calculate the correlation between the scores provided by each metric for every hypothesis-reference pair against their corresponding 
$GoldCER$ values. 
This evaluation provides a fine-grained assessment demonstrating the ability of each metric to distinguish the amount of errors in each hypothesis. 
In the system-level evaluation, we calculate the overall $GoldCER$ for each of the three systems (\textit{HMM-DNN}, \textit{Conformer-A}, and \textit{Conformer-F}) which acts as the ground truth score. We then obtain the overall scores for the three systems using each evaluation metric, assessing its ability to provide correct system ranking.

\subsection{Results and Discussion}

\subsubsection{Overall Sentence-level Evaluation}

\begin{table*}[th!]
\centering
 \includegraphics[width=0.9\textwidth]{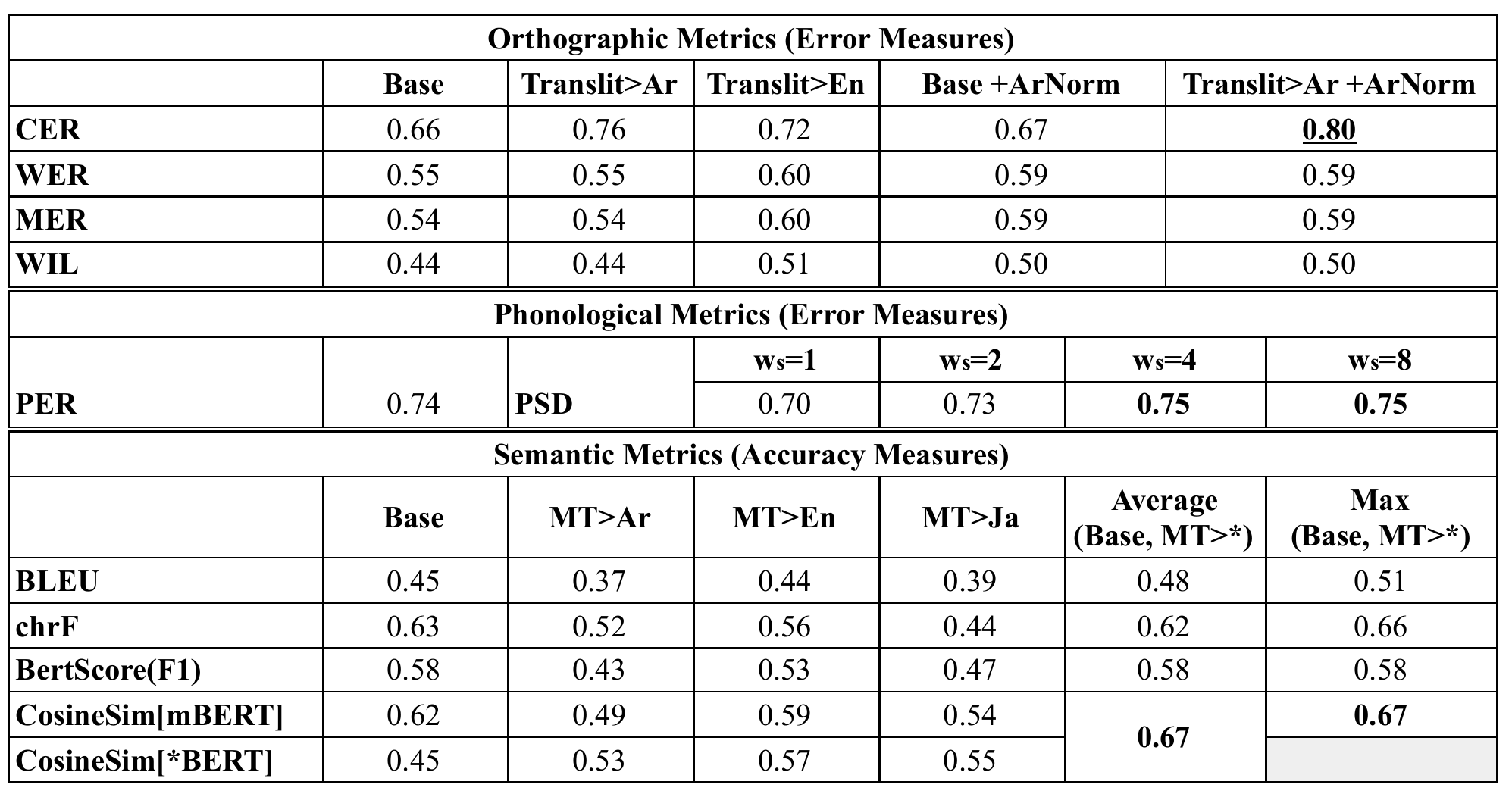}
    \caption{Sentence-level correlations calculated between $GoldCER$ and the scores of different metrics. Given that the orthographic and phonological evaluation metrics used are error metrics, while the semantic metrics are accuracy metrics, we present the correlations against (1-$GoldCER$) for semantic metrics, to consistently report positive correlations for easier readability. 
    CosineSim[mBERT] are the results achieved using mBERT. CosineSim[*BERT] are the results achieved using BiBERT for the original hypothesis-reference pairs (Base), CAMeLBERT for Arabic translations (which outperformed AraBERT), BERT-base for English translations, and BERT-base-japanese for Japanese (which outperformed BERT-large-japanese).}
\label{table:results_sentenceLevel}
 \vspace{-0.2cm}
\end{table*}

In Table \ref{table:results_sentenceLevel}, we present the sentence-level correlations between $GoldCER$ and different metrics' scores. 
For the \textbf{orthographic metrics}, 
we demonstrate the correlations using CER, WER, MER, and WIL applied on the original hypothesis-reference pairs (Base) as well as their transliterations into Arabic (\textit{Translit$>$Ar}) and Roman (\textit{Translit$>$En}) scripts. We also show the effect of applying Alif/Ya normalization on the original sentences (\textit{Base+ArNorm}) and the transliterations in Arabic script (Translit$>$Ar+ArNorm). 
We observe that across the different settings, 
the highest correlations are achieved using CER, followed by WER and MER, then WIL. Similar to the findings in \cite{AKH19}, with normalization, higher correlations are achieved, which we report for all the four metrics. When applying CER, transliterating to Arabic outperforms transliterating to English. However, for word-level evaluation metrics (WER, MER, and WIL), transliterating to English gives higher correlations, even though the references are dominated by Arabic words (77\%). 
This can be justified by the ability to resolve Arabic unstandardized orthography issues when transliterating into English. 

\begin{table}[th!]
\centering
 \includegraphics[width=\columnwidth]{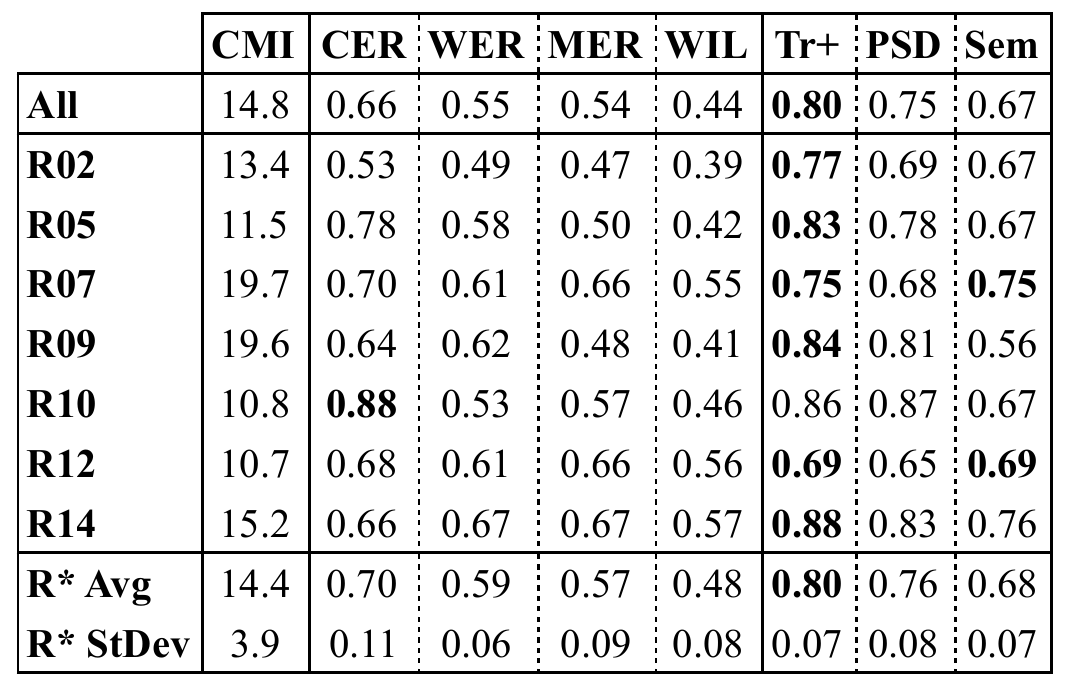}
 \caption{Reported sentence-level correlations per recording, with recording-level Code-Mixing Index (CMI in percentage). We show the correlations for CER, WER, MER, WIL, in addition to transliterating to Arabic followed by Alif/Ya normalization (Tr+), PSD ($w_{s}=4$), and $Average(Base, MT>*)$ for the semantic measure (Sem).}
 \label{table:results_sentenceLevelPerSpeaker}
 \vspace{-0.3cm}
\end{table}

\begin{table*}[th!]
\centering
 \includegraphics[width=\textwidth]{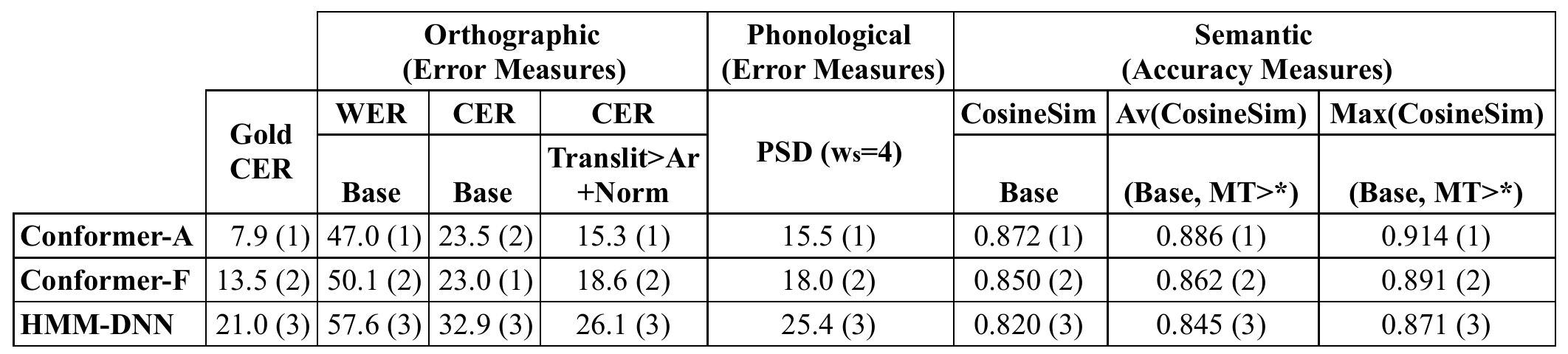}
 \caption{System-level overall scores and ranking (denoted between parentheses) across the different evaluation metrics.}
 \label{table:results_systemLevel}
 \vspace{-0.3cm}
\end{table*}

In the scope of \textbf{phonological metrics}, we 
find that $w_S = 4$ provides the highest correlation of PSD with 
human judgment. The efficacy of incorporating phonological similarity in the error calculation is demonstrated, where PSD outperforms PER (Phoneme Error Rate). For \textbf{semantic metrics}, the highest correlations are achieved using cosine similarity, followed by chrF, BertScore(F1), then BLEU. 
Across the different metrics, higher correlations are achieved by applying the metrics directly on the original text rather than on translated text. This can be foreseen, as the success of this approach is dependent on the performance of the underlying MT system. By looking into the translations, it is obvious that a significant amount of translation errors is introduced, 
which is propagated to the metric scores. 
However, the translation step still proves to be beneficial, where across the metrics, higher correlations are achieved by applying sentence-level aggregation to the scores 
achieved across the different language setups, where we have tried $Avg$ and $Max$ functions.\footnote{For $Avg$, the score for each reference-hypothesis pair is calculated as the average of cosine similarity scores achieved across the four language setups, where we choose the best-performing pretrained model for each language. For $Max$, the score for each pair is calculated as the maximum cosine similarity score across the four language setups using mBERT model. We also investigate taking the average of cosine similarity scores using mBERT model, however, it gives lower correlations.}
The highest correlation for semantic metrics is achieved by aggregating the cosine similarity scores using the $Avg$ function across the original sentences and the three translations. 
While the semantic-based metrics provide lower correlations compared to transliteration and phonetic similarity, we believe that the efficacy of the proposed translation pipeline should be revisited with future advances in CS MT systems.

In general, results show that transliteration and phonetic similarity outperform conventional CER, WER, MER, and WIL evaluation metrics applied directly on the original sentences. Despite the limitations of the semantic measures, it outperforms WER, MER, and WIL, and performs equally well as CER. The highest correlation is achieved by using CER over the texts transliterated into Arabic script followed by Arabic normalization.

\subsubsection{Recording-based Sentence-level Evaluation}
To further investigate the validity of our results, we 
perform the same sentence-level evaluation 
across the seven different recordings in our corpus, as presented in Table \ref{table:results_sentenceLevelPerSpeaker}. Given that the recordings contain different degrees of CS, this analysis allows us to evaluate the consistency of our results. 
We measure CS levels in terms of \textit{Code-Mixing Index} (CMI) \cite{GD16,chowdhury2020effects}, calculated on the utterance-level as follows:
\[C_{u}(x)=100 * \frac{\frac{1}{2}*(N(x)-max_{L_i\in \textbf{L}}\{t_{L_i}\}(x))+\frac{1}{2}P(x)}{N(x)} \]

where $N$ is the number of language-dependent tokens in utterance $x$; $L_i \in \textbf{L}$ the set
of all languages in the corpus; $max\{t_{L_i}\}$ represents the number of tokens in the dominating language in $x$, with $1 \leq max\{t_{L_i}\}$$ \leq N$; and $P$ is the number of code alternation points in $x$; $0 \leq P < N$. We calculate the recording-level CMI by averaging the utterance-level values. We note that there is a degree of variety in CMI values across recordings ranging from 11 to 20. 
In Table \ref{table:results_sentenceLevelPerSpeaker}, we show the recording-level CMI values 
and the correlations for CER, WER, MER, WIL, transliterating to Arabic followed by Alif/Ya normalization (Tr+), PSD ($w_{s}=4$), and $Average(Base, MT>*)$ for the semantic measure (Sem). We observe that for the conventional metrics, it is mostly the case that CER outperforms WER/MER (performing on-par), which outperform WIL. We confirm that transliterating to Arabic followed by Alif/Ya normalization is the best-performing metric, outperforming CER for 6/7 recordings. In the case where CER achieved highest correlation, it is only slightly better than transliteration. 
By looking into the standard deviation of the metrics' correlation scores, 
we observe that CER has the highest value (0.11), compared to transliteration (0.07), phonological (0.08) and semantic (0.07) measures. This reflects that CER's performance is less consistent than other metrics. 
In future work, we plan to understand the relation between the correlation scores and CS behaviour as well as other variables. 

\subsubsection{System-level Evaluation}
Results for the system-level evaluation are presented in Table \ref{table:results_systemLevel}. For semantic-based metrics, the overall score is calculated as the average of sentence-level scores. As indicated by the $GoldCER$, the ranking of the systems is: \textit{Conformer-A}, \textit{Conformer-F}, and \textit{HMM-DNN}. We show that, apart from CER applied directly on the original sentences, all the evaluation metrics provide the same ranking conclusion. The relative scores of the systems are not equivalently reflected in all metrics though, which is worth further investigations. In the future, we plan to include more systems, in order to be able to derive correlations between systems' overall scores.


\section{Conclusion and Future Work} 

In this work, we \textit{(i)} develop a corpus of human judgment -- with minimal edits of different ASR output; and \textit{(ii)} benchmark the performance of different evaluation metrics and their ability to correctly evaluate CS ASR outputs in correlation to the ground truth human post-editing effort. 
We cover commonly-used evaluation metrics, in addition to three approaches aiming at handling CS challenges: transliteration, phonetic similarity, and semantic similarity. Our results show that WER and CER are not adequate for evaluating CS languages having cross-transcription and spelling variation. 
The highest correlation to the post-editing effort is achieved by transliteration followed by phonetic similarity, semantic similarity, CER, and WER, in order. 
In future, we plan to evaluate the proposed methods for the MUCS2021 \cite{diwan2021multilingual} challenge to ensure generalization across more languages. Furthermore, we plan to create the human acceptability corpus for language pairs sharing the same writing script.
\section{ACKNOWLEDGMENTS}
The work presented here was carried out during the 2022 Jelinek Memorial Summer Workshop on Speech and Language Technologies at Johns Hopkins University, which was supported with funding from Amazon, Microsoft and Google. We also thank the anonymous reviewers for their helpful feedback.


\bibliographystyle{IEEEbib}
\bibliography{main}

\end{document}